\documentclass[prl,twocolumn,amsmath,amssymb,superscriptaddress]{revtex4-1}
\usepackage{graphicx}
\usepackage{ulem}
\usepackage{graphicx}
\usepackage{dcolumn}
\usepackage{bm}
\usepackage{mathrsfs}
\usepackage{subfigure}
\usepackage{natbib}
\usepackage{color}
\usepackage{amssymb}
\usepackage{amsmath}
\usepackage{bbm}

\begin{document}

\title{Shallow-band superconductors: pushing superconductivity types apart}

\author{S. Wolf}
\affiliation{Institut f\"{u}r Theoretische Physik III, Bayreuth
Universit\"{a}t, Bayreuth, Germany}

\author{A. Vagov}
\affiliation{Institut f\"{u}r Theoretische Physik III, Bayreuth
Universit\"{a}t, Bayreuth, Germany}

\author{A. A. Shanenko}
\affiliation{Departamento de F\'isica, Universidade Federal de
Pernambuco, Recife, PE, Brazil}

\author{V. M. Axt}
\affiliation{Institut f\"{u}r Theoretische Physik III, Bayreuth
Universit\"{a}t, Bayreuth, Germany}

\author{J. Albino Aguiar}
\affiliation{Departamento de F\'isica, Universidade Federal de
Pernambuco, Recife, PE, Brazil}

\begin{abstract}
Magnetic response is a fundamental property of superconducting materials,
which helps to distinguish two superconductivity types: the 
ideally diamagnetic type I and type II, which can develop
the mixed state with Abrikosov vortices. We demonstrate that multi-band
superconductors with one shallow band, that have recently attracted much
attention for their high critical temperatures and unusual properties, often stay 
apart of this simple 
classification. In a wide range of  microscopic parameters such systems 
fall into the inter-type or transitional interval in between the standard types 
and display unconventional mixed state configurations.
\end{abstract}


\date{\today}

\maketitle

The ongoing search for superconductors 
with higher transition temperatures ignited 
intense research on materials with many carrier bands 
(multi-band systems) with so-called shallow bands~\cite{boz}. 
It has long been known that superconductivity properties in multi-band materials are very
sensitive to the position of the chemical potential.  As was first predicted in 1970's if a chemical potential touches the lowest energy point of a band with a large density of states (DOS) the superconductivity may enter the regime of the BCS-BEC crossover \cite{eagl}. Evidences of this regime have been recently observed in multi-band  ${\rm Fe Se}_x{\rm Te}_{1-x}$ \cite{kan}. However, the research on the role of shallow bands is, at present, limited to a few microscopic characteristics  of the BCS-BEC crossover, such as increased fluctuations and the appearance of the pseudogap  (for recent developments see \cite{dal,str} and references therein). Until now, manifestations of the BCS-BEC crossover on a macroscopic level, 
in particular possible modifications of the magnetic properties, remain largely unknown. 
In this Letter we demonstrate that many multi-band superconductors 
with a shallow band are in a mixed state and reveal
magnetic properties that cannot be attributed 
to standard superconductivity types.

\begin{figure*}[t]
\begin{center}
\resizebox{1.7\columnwidth}{!}{\rotatebox{0}{
\includegraphics{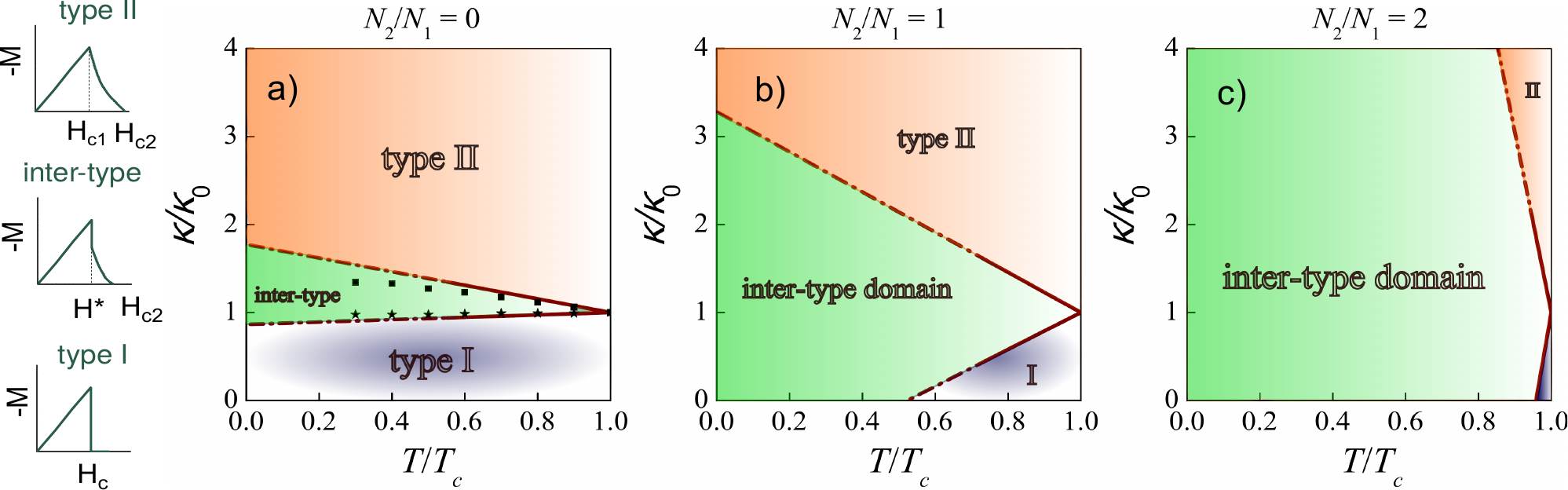}}}
\end{center}
\vspace{-0.4cm}
\caption{Phase diagram of a two-band superconductor on the $(\kappa,T)$-plane. The left panels illustrates schematically the magnetization field dependence of types I and II and of the inter-type domain. Panels a), b) and c) 
correspond to $\eta = 0$, $1$ and $2$, respectively, demonstrating progressive
widening of the inter-type domain. For comparison, dots in the left panel represent numerical results for the inter-type boundaries (squares for $\kappa_{max}^\ast = \kappa_{li}^\ast $ and stars for $\kappa_{min}^\ast = \kappa_{2}^\ast $), obtained by solving the Eilenberger equation \cite{miran}. }
\vspace{-0.4cm}
\label{fig1}
\end{figure*}

We start by recalling the well known fact that the Ginzburg-Landau (GL) theory
predicts that the magnetic response of a superconductor is determined by the GL
parameter $\kappa=\lambda/\xi$, where $\lambda$ is the magnetic penetration
depth and $\xi$ is the GL coherence length ~\cite{degen,landau9,kett}. The
superconductivity types interchange sharply at the critical GL parameter
$\kappa_0=1/\sqrt{2}$, so that at $\kappa <  \kappa_0$ or $\kappa > \kappa_0$ a
superconductor belongs,  respectively, to type I or II. More elaborate
theoretical analysis~\cite{jacobs1} as well as experimental
studies~\cite{krag,ess,ast,auer} have led to the conclusion that the type
interchange takes place over a finite interval of $\kappa$'s, here referred to as
{\it inter-type interval} or  {\it inter-type domain}, if considered in the
entire $(\kappa,T)$-plane.  In conventional single-band materials this domain is
narrow and is usually ignored in most discussions of the superconductivity
types. Superconductivity properties in this interval are investigated in detail
neither theoretically nor experimentally. The studies presented so far
suggest that it has many non-conventional phenomena, not found in standard bulk
superconductors. In particular, unlike type II materials such inter-type superconductors demonstrate the first order first-order phase transition between the Meissner and the mixed states, which is associated with non-monotonic vortex-vortex
interactions~\cite{jacobs2} (schematical magnetization curves are shown in Fig. \ref{fig1}).  
They also demonstrate stable multi-quantum vortices~\cite{lasher,ovch} and
a giant paramagnetic Meissner effect~\cite{silva}. 

In this work,  using a two-band
prototype model of a superconductor, we demonstrate that the presence of a
shallow band with strong pairing interaction dramatically enlarges the inter-type
domain and thus changes qualitatively the magnetic properties  of such systems.
To this end we study details of the interchange of  superconductivity types. A
standard description of the interchange in the GL theory is done by introducing
a critical GL parameter $\kappa^{\ast}$, which marks the appearance of the  mixed
state at the thermodynamic critical field $H_c$, so that at  $\kappa >
\kappa^{\ast}$ it wins energetically over the uniform Meissner state. It is
obtained by solving the equation:
\begin{align}
\mathfrak{G}(\kappa^{\ast},T) = 0, \quad \mathfrak{G} =\int\mathfrak{g}\, d{\bf r} , \quad \mathfrak{g} =
\mathfrak{f} +\frac{H^2_c}{8\pi}- \frac{H_c B}{4\pi} \label{eq:critical_parameter}
\end{align}
where $\mathfrak{G}$ is the Gibbs energy difference  between the mixed and the Meissner states calculated at $H_c$ and  $\mathfrak{f}$ is the condensate free-energy density. 
The magnetic induction ${\bf B}$ is assumed parallel to the external field $H=H_c$. 

In the GL theory a particular choice of the non-uniform mixed state is not
important for criterion (\ref{eq:critical_parameter}): one obtains the same
critical parameter $\kappa^\ast = \kappa_0$ for all possible flux
configurations. The reason is a special degeneracy of the GL equations at
$\kappa_0$, often referred to as the Bogomolny point \cite{bogomol}. However,
beyond the GL theory at $T< T_c$ this degeneracy is removed and
$\kappa^{\ast}=\kappa_i^{\ast}$ depends on the flux configuration $i$. 
The number of topologically
different configurations $i$ is infinite and so is the number of critical
parameters $\kappa_i^{\ast}$. This defines a finite inter-type interval
$[\kappa^\ast_{min},\kappa^\ast_{max}]$, where superconductivity types
interchange gradually by a sequential appearance of flux configurations $i$ when
$\kappa_i^{\ast}$ is crossed. Previous research indicates that the lower
boundary of this interval is defined by the start of superconductivity
nucleation at $H_c$ (this is equivalent to the condition $H_{c2} = H_c$), while
the upper boundary if determined by the appearance of a long-range attraction
between two Abrikosov vortices.


We now calculate the inter-type boundaries for a model with two carrier bands,
one of which is a shallow 2D band with a  minimal energy that coincides with the
chemical potential. The lower dimensionality of the band increases its DOS, which
is crucial for reaching the BCS-BEC crossover regime \cite{shanenko}. The
other band is deep. Its
dimensionality  is not so important, although a 2D model
requires more attention in view of thermal fluctuations. However, once it is
assumed that the superconductivity in the deep band can be described within the
mean field approach, details of the model are not critical. For simplicity we
assume that both bands are two dimensional.  

According to the BCS theory the free-energy density of the condensate state 
in a two-band system writes as:
\begin{align}
\mathfrak{f}=\frac{{\bf
B}^2}{8\pi}+\Delta^{\dagger}\check{g}^{-1}\Delta +
\sum\limits_{\nu=1,2}\mathfrak{f}_{\nu}, \label{eq:f}
\end{align} where $\Delta^{\dagger} = \big(\Delta^\ast_1({\bf r}),
\Delta^\ast_2({\bf r})\big)$ is the band gap function and $\check{g}^{-1}$ is
the inverted $2\times2$ coupling matrix with real $g_{ij} = g_{ji}$. We
calculate the free energy using the perturbation expansion with the small
parameter $\tau = 1 - T/T_c$, which leads to the so-called extended GL (EGL)
theory ~\cite{extGL1,extGL2}. For a two-band system the lowest order of the
expansion yields the standard GL theory with a single order parameter
\cite{extGL2,kogan}. A far-reaching consequence of this fact is that two-band
superconductors follow the standard classification, where types I and II are
separated by the Bogomolny point at $T_c$ and by a finite inter-type interval at
$T<T_c$.   

The perturbation expansion is derived similarly to Ref. \cite{extGL2}, with the
difference that here the bands are qualitatively different (shallow and deep),
which leads to additional contributions to the series expansion. Here, we
presents a sketch of the derivation which highlights the differences with
Ref. \cite{extGL2}. Expanding $\mathfrak{f}_{\nu}$ in Eq.~(\ref{eq:f}) in
powers of $\Delta_\nu$ and its gradients yields:
\begin{align}
\mathfrak{f}_{\nu} = &- a_{1,\nu} |\Delta_\nu|^2 + a_{2,\nu}
|{\bf D}\Delta_\nu|^2 - a_{3,\nu}\Big(|{\bf D}^2
\Delta_\nu|^2 \notag \\
& +\frac{{\rm rot}{\bf B}\cdot{\bf i}_{\nu}}{3}+ \frac{4e^2}{\hbar^2
\mathbbm{c}^2}{\bf B}^2 |\Delta_\nu|^2\Big)
+ a_{4,\nu}{\bf B}^2|\Delta_\nu|^2 \notag \\
& +\frac{b_{1,\nu}}{2}|\Delta_\nu|^4 -\frac{b_{2,\nu}}{2}
\Big(L_{\nu} |\Delta_\nu|^2|{\bf D}\Delta_\nu|^2 \notag\\
&+l_{\nu} \big[(\Delta_\nu^{\ast})^2 ({\bf D}\Delta_\nu)^2 + {\rm
c.c.}\big]\Big)- \frac{c_{1,\nu}}{3}|\Delta_\nu|^6, 
\label{eq:f_nu}
\end{align}
where the band coefficients $a_{n,\nu}$, $b_{n,\nu}$, $c_{n,\nu}$ are $T$-dependent. 
The constants $L_{\nu}$ and $l_{\nu}$ are introduced to capture the differences 
between the bands and 
\begin{align}
{\bf i}_{\nu} = \frac{4 e}{\hbar\,\mathbbm{c}}{\rm Im} \big[ \Delta_{\nu}{\bf D}^\ast \,\Delta_{\nu}^\ast \big], \quad \boldsymbol{D}=\boldsymbol{\nabla}- \frac{\mathbbm{i} 2e}
{\hbar\,\mathbbm{c}} \boldsymbol{A}.  
\label{eq:i_nu}
\end{align}
The $\tau$-expansion is obtained by representing all quantities in 
Eq.~(\ref{eq:f_nu}) as $\tau$-series:  
\begin{align}
&\Delta_{\nu} =\tau^{1/2} \big[\Delta_{\nu}^{(0)} + \tau \Delta_{\nu}^{(1)}\big],\quad {\bf A} = \tau^{1/2} \big[{\bf A}^{(0)} + \tau {\bf A}^{(1)}\big], \notag\\
& {\bf B} = \tau \big[{\bf B}^{(0)} + \tau {\bf B}^{(1)} \big], \quad H_c = \tau \big[H^{(0)}_c+\tau H^{(1)}_c \big],
\label{eq:DelBA}
\end{align} 
where the two lowest orders, needed to derive the leading order
corrections to the GL theory, are kept. We take into account the $\tau$-scaling of
the coordinates \cite{extGL1}, which introduces an additional factor $\sqrt{\tau}$
for each gradient in the series. Expanding the temperature-dependent
coefficients in Eq.~(\ref{eq:f_nu}) yields:
\begin{align}
&a_{1,\nu} = {\cal A}_{\nu} - \tau\big[a^{(0)}_{\nu} +
\tau a^{(1)}_{\nu}\big],\quad a_{2,\nu} = {\cal K}^{(0)}_{\nu} + \tau{\cal K}^{(1)}_{\nu},\notag\\
&a_{3,\nu}={\cal Q}^{(0)}_{\nu},\quad a_{4,\nu} =
r^{(0)}_{\nu},\quad b_{1,\nu} = b^{(0)}_{\nu}+\tau b^{(1)}_{\nu},\notag \\
&b_{2,\nu}L_{\nu}= {\cal L}^{(0)}_{\nu},\quad b_{2,\nu}l_{\nu}= \ell^{(0)}_{\nu},\quad c_{1,\nu}=
c^{(0)}_{\nu}, 
\label{eq:coef_exp}
\end{align} 
where the coefficients are calculated from the chosen microscopic
model for the band states. Substituting Eqs.~(\ref{eq:DelBA}),
(\ref{eq:coef_exp}) and the gradient scaling into Eq.~(\ref{eq:f_nu}) and then
calculating the integrals in Eq.~(\ref{eq:critical_parameter}) one derives the
$\tau$-expansion for $\mathfrak{G}$.  

\begin{table*}
\begin{ruledtabular}
\begin{tabular}{ccccccccc}
{\color{black} $\nu$} & ${\cal M}^{(0)}_{b,\nu}$ & ${\cal
M}^{(0)}_{c,\nu}$ & ${\cal M}^{(0)}_{{\cal K},\nu}$ & ${\cal
M}^{(0)}_{{\cal Q},\nu}$ & ${\cal M}^{(0)}_{{\cal L},\nu}$ & ${\cal
M}^{(1)}_{a,\nu}$ & ${\cal M}^{(1)}_{b,\nu}$ & ${\cal
M}^{(1)}_{{\cal K},\nu}$\\[1ex]
\hline
\\
1 & $7\zeta(3)/(8\pi^2)$ & $93\zeta(5)/(128\pi^4)$ &
$7\zeta(3)/(32\pi^2)$ & $93\zeta(5)/(2048\pi^4)$ &
$31\zeta(5)/(32\pi^4)$ & 1/2 & 2 & 2\\[1ex]
2 & $7\zeta(3)/(8\pi^2)$ & $93\zeta(5)/(128\pi^4)$ &
$3\zeta(2)/(8\pi^2)$  & $7\zeta(3)/(512\pi^2)$ &
$25\zeta(4)/(16\pi^4)$ & 1/2
& 2 & 1\\[1ex]
\end{tabular}
\caption{\label{tab:tab1} Numerical factors ${\cal
M}^{(0)}_{w,\nu}$~(with $w=b,c,{\cal K},{\cal Q},{\cal K}$) and
${\cal M}^{(1)}_{w,\nu}$~(with $w=a,b,{\cal K}$) for the deep
($\nu=1$) and shallow ($\nu=2$) bands, see Eqs.~(\ref{eq:c_kappa4A})
and (\ref{eq:c_kappa4B}). In the table $\zeta(x)$ is the Riemann
zeta function of $x$.}
\end{ruledtabular}
\end{table*}

It is important that the leading correction to the GL free energy requires only
$\Delta_{1,2}^{(0)}$ and ${\bf B}^{(0)}$ while $\Delta_{1,2}^{(1)}$ and ${\bf
B}^{(1)}$ are not needed \cite{extGL2}. 
Thus, the corrected free energy can be evaluated
from the knowledge of the solution of the GL equations alone.
The latter exhibits a single order parameter $\Psi$, which determines both 
gaps by:
\begin{align}
 \left(
\begin{array}{c}
\Delta_1^{(0)}\\
\Delta_2^{(0)}
\end{array}
\right) = \left(
\begin{array}{c}
S^{-1/2}\\
S^{1/2}
\end{array}
\right) \Psi({\bf r}). \label{eq:Psi}
\end{align}
The band weight factor $S$ is obtained by solving the 
linearized gap equation for $T_c$ which yields:
\begin{align}
S = \frac{1}{g_{12}} \big(g_{22} - G {\cal A}_1\big)=
\frac{g_{12}}{g_{11}-G{\cal A}_2},
\label{eq:S}
\end{align}
where $G= {\rm det}[ g] =  g_{11}g_{22} - g_{12}^2$. The integration in 
Eq.~(\ref{eq:critical_parameter}) is simplified with the help of 
the GL equations and the final result for $\mathfrak{G}$ depends only on 
the integrals: 
\begin{align}
{\cal I} =\!\! \int |\Psi|^2 \big(1 - |\Psi|^2\big)d{\bf r},\;\,\,{\cal
J} =\!\! \int |\Psi|^4 \big(1 - |\Psi|^2\big)d{\bf r}.
\label{eq:I_J}
\end{align}
Solving Eq.~(\ref{eq:critical_parameter}) up to the leading order 
corrections of the GL theory we obtain: 
\begin{align}
\kappa^\ast = \kappa_0 +\tau \kappa^{\ast (1)}
\label{eq:kappa}
\end{align}
with
\begin{align}
\frac{\kappa^{\ast(1)}}{\kappa_0} =&\;\bar{\cal K} -\bar{c}
+2\bar{\cal Q}+\bar{G}\,\bar{\beta}\big(2\bar{\alpha}
-\bar{\beta}\big)\notag\\
&+\frac{{\cal J}}{\cal I}\left(\frac{\bar{\cal L}}{4} - \bar{c}
-\frac{5}{3}\bar{\cal Q} - \bar{G}\bar{\beta}^2\right),
\label{eq:kappa1}
\end{align}
where the dimensionless constants read as:
\begin{align}
&\bar{\cal K} = \frac{{\cal K}^{(1)}}{\cal K}-\frac{{b}^{(1)}}{2b},
\quad \bar c =\frac{c a }{3 b^2},\quad \bar {\cal Q}=\frac{a{\cal Q}}{{\cal
K}^2},\quad \bar{\cal L}=\frac{a {\cal L}}{b{\cal K}},\notag\\
&\bar{G} =\frac{G a }{4g_{12}}, \quad \bar
\alpha=\frac{\alpha}{a}-\frac{{\Gamma}}{{\cal K}}, \quad  \bar\beta =
\frac{\beta}{b}-\frac{{\Gamma}}{{\cal K}}.
\label{eq:c_kappa1}
\end{align}
The coefficients are defined by the band contributions as: 
\begin{align}
&\omega = \frac{\omega_1^{(0)}}{S^{p}} +S^p w_2^{(0)}, 
\quad \omega^{(1)}=\frac{\omega_1^{(1)}}{S^{p}} + S^p w_2^{(1)},  \notag \\
&\alpha=\frac{a_1^{(0)}}{S} - S a_2^{(0)},\;
\beta=\frac{b_1^{(0)}}{S^{2}} -S^2 b_2^{(0)},\notag \\
&\Gamma = \frac{{\cal K}_1^{(0)}}{S} -S {\cal K}_2^{(0)}.
\label{eq:c_kappa2}
\end{align}
where $\omega = \{a,{\cal K},{\cal Q},r,b,{\cal L},c\}$, $\omega^{(1)} = \{ {\cal K}^{(1)}, b^{(1)}\}$, 
$w_\nu^{(0)} = \{ a_\nu^{(0)},{\cal K}_\nu^{(0)},{\cal Q}_\nu^{(0)},r_\nu^{(0)},b_\nu^{(0)},{\cal L}_\nu^{(0)},c_\nu^{(0)}\}$ 
and values $p=\{1,2,3\}$ appear respectively for coefficients $a_{n,\nu}$, $b_{n,\nu}$ and $c_{n,\nu}$. 

The band coefficients in Eq.~(\ref{eq:coef_exp}) are calculated for 
a model with 2D quadratic dispersion for both bands. We note that in the EGL formalism
the band dimensionality affects the results mainly via the ratio between the
band DOSs. The calculations are done in the clean limit. For the deep band
($\nu =1$) the inequality $\Delta_{1} \ll \mu - \varepsilon_{1,k=0}$ means that we
can use standard approximations employed in the derivations of the EGL theory
for the 3D case \cite{extGL1}. For the shallow band ($\nu = 2$) the chemical
potential is assumed to coincide with the band minimum, $\mu =
\varepsilon_{2,k=0}$. Then the leading order coefficients in
Eq.~(\ref{eq:coef_exp}) are:   
\begin{align}
& {\cal A}_\nu = N_\nu \ln\Big(\frac{2e^\gamma\hbar \omega_c}{\pi
T_c}\Big),\, a^{(0)}_{\nu} = - N_{\nu},\,b^{(0)}_{\nu} = N_{\nu}
\frac{{\cal M}^{(0)}_{b,\nu}}{T_c^2},\notag\\
&c^{(0)}_{\nu} =N_{\nu}\frac{{\cal M}^{(0)}_{c,\nu}}{T_c^4},\, {\cal
K}^{(0)}_{\nu} = N_{\nu}{\cal M}^{(0)}_{\cal
K,\nu}\frac{\hbar^2v_{\nu}^2}{T_c^2},\notag\\
&{\cal Q}^{(0)}_{\nu} = N_{\nu} {\cal M}^{(0)}_{{\cal Q},{\nu}}
\frac{\hbar^4 v_{{\nu}}^4}{T_c^4},\;{\cal L}^{(0)}_{\nu} =
N_{\nu}\,{\cal M}^{(0)}_{{\cal L},\nu}\frac{\hbar^2
v_{\nu}^2}{T_c^4}, 
\label{eq:c_kappa4A}
\end{align}
where $\hbar\omega_c$ is the cut-off energy, $\gamma$ is the Euler constant, $N_{\nu}$
is the band DOS, $v_{\nu}$ denotes the characteristic band
velocity, i.e.,  the Fermi velocity $v_F = \sqrt{2\mu/m_\nu}$ for the
deep band and the temperature velocity $v_T
=\sqrt{2T_c/m_\nu}$ for the shallow band. 
The additional numerical factors ${\cal M}^{(0,1)}_{w,\nu}$ are listed in
Tab.~\ref{tab:tab1}. The band DOSs are $N_\nu = \tilde{N}_{\nu}
m_\nu/(2\pi\hbar^2)$ with $\tilde{N}_\nu$ being an additional factor that accounts for the density of states in z-direction 
(this quantity accounts for the 3D character of the entire system and does not affect the final conclusions). 
The next-order coefficients in Eq.~(\ref{eq:coef_exp}) are given by:
\begin{align}
w^{(1)}_{\nu}={\cal M}^{(1)}_{w,\nu}w^{(0)}_{\nu},
\label{eq:c_kappa4B}
\end{align}
where $w=\{a,{\cal K},b\}$. 


Knowing the ratio $\eta = N_2/N_1$ of the DOSs and the coupling constants
$\lambda_{ij}=g_{ij}N$~($N=N_1+N_2$) one obtains the critical temperature $T_c$
from the linearized gap equation and then $\kappa^{\ast}$  from
Eq.~(\ref{eq:kappa1}). It is important that apart from $T_c$ the final
expression for $\kappa^{\ast}$ depends only on $\eta$ and $v_2/v_1$, but not on
$N_{1,2}$ and $v_{1,2}$ separately. The ratio ${\cal J}/{\cal I}$ in
Eq.~(\ref{eq:kappa1}) is calculated using the solution of the GL equations at
$\kappa_0$, which at this point reduces to the pair of self-dual Sarma-Bogomolny
equations \cite{degen,bogomol}. 

As mentioned before, the lowest boundary of the inter-type interval
$\kappa^\ast_{min}$ is calculated from the condition that the inhomogeneous
mixed state disappears at $H_c$. It follows from Eq.~(\ref{eq:I_J}) that in the
limit of  a vanishing mixed state one has ${\cal J}/{\cal I} = 0$.  The upper
boundary $\kappa^\ast_{max}$ is defined by the condition that the sign in the
long-range vortex-vortex interaction changes. In this case we need to calculate
the long-range asymptote of ${\cal J}(R)/{\cal I}(R)$  for the two-vortex
solution as a function of the distance $R$ between the vortices.  This can be done
analytically yielding the exact asymptote ${\cal J}(R)/{\cal I}(R) =2$ at $R\to
\infty$. 

Although the final expression for $\kappa^\ast$ is a complicated algebraic
function of the  microscopic model parameters, it can be considerably simplified
for the case of a two-band system with $v_2/v_1  \sim \sqrt{T_c / \mu} \ll 1$ 
in the vicinity of the BCS-BEC crossover, where the contribution of the shallow
band to the condensate state is dominant and thus $S \gtrsim 1$ in
Eq.~(\ref{eq:S}) \cite{shanenko}. We obtain:
\begin{align}
 \frac{\kappa^{ (1)\ast}}{\kappa_0} \approx   \tilde {\cal Q} \left( 2 - \frac{5}{3} \frac{\cal J}{\cal I}  \right) S^2  \eta , \quad \tilde {\cal Q} = \frac{a_1^{(0)} {\cal Q}_1^{(0)}}{{\cal K}_1^{(0)2}},
 \label{eq:kappa_approx}
\end{align} where the bracket gives 2 for $\kappa_2^\ast$ and $-4/3$ for
$\kappa_{li}^\ast$, if we use the above results for the  ${\cal J}/{\cal I}$
ratio. According to this simplified expression the width of the inter-type domain
is governed by the ratio of the DOSs $\eta$, the couplings (via $S$)
and the dimensionless constant $\tilde {\cal Q}$. Equations (\ref{eq:f_nu}) and
(\ref{eq:coef_exp}) show that $\tilde {\cal Q}$  controls the contribution of the
fourth-order gradient term in the free-energy expansion for the deep band. 

Taking into account that $S$ increases with $\eta$, when the system is close to
the crossover \cite{shanenko}, one concludes that the inter-type interval
increases with $\eta$. This widening of the inter-type domain in the
$(\kappa,T)$-plane is illustrated in Fig.~\ref{fig1}, where three panels show
$\kappa_2^\ast$ and  $\kappa_{li}^\ast$ as  functions of temperature (this is of
cause a linear dependence in the EGL theory) calculated at $\eta = 0,1,2$ using
the non-simplified expressions for $\kappa^\ast$. We note that 
$\kappa^\ast$ depends only modestly on the coupling constants  but is indeed very
sensitive to the value of $\eta$. We also note that when $\eta = 0$ (left panel) only
the deep band is involved in the condensate formation so that our results should
be comparable with those obtained earlier for a  single-band model. Indeed, a comparison
with microscopic numerical calculations for a 2D system \cite{miran},
shown by dots in the left panel of Fig.~\ref{fig1}, reveals a very good
quantitative agreement down to temperatures $0.5T_c$. At larger $\eta$ the
contribution of the shallow band  to the condensate  increases and the
inter-type domain widens sharply, as shown in the middle and right panels of
Fig.~\ref{fig1}. 

We now address the question whether thermal fluctuations invalidate the mean
field foundations of the GL and EGL approaches. It is well known that the existence
of shallow bands strongly enhances fluctuations. This can be seen by
considering the fluctuation-related contribution to the heat capacity. The GL
theory for a single-band system yields for this quantity $\delta C_V \sim
(L/\xi_0)^2 \tau^{-1} $, where $L$ is the sample length and $\xi_0^2 = -{\cal K}
/a$ is the zero-temperature GL coherence length \cite{kett}. Here, this length is
determined by the coefficients of the GL equation, but it is also related to the
microscopic BCS coherence length or the Cooper-pair size. One notices that the
coherence length calculated separately for the shallow band, $\xi_{0,2} \sim
v_2$, is rather small and, therefore, the corresponding Ginzburg-Levanjuk
parameter $Gi_2$ (temperature interval around $T_c$, where the fluctuations in
this band are important) may become comparable with the temperature interval
where the EGL theory can be used.  However, in a two-band system the
fluctuations are screened due to the interactions with the deep band. This
follows from the calculations of the full two-band coherence length, which
yields $\xi_0^2 = \sum_\nu \rho_\nu\xi_{0,\nu}^2$ with $\rho_\nu$ being the band
weight factors. The large coherence length of the deep band, $\xi_{0,1}$, ensures
that $\xi_0$ is not very small (unless $\rho_1 \ll \rho_2$). In the same limit,
that was used to derive Eq.~(\ref{eq:kappa_approx}), one obtains $\xi_0 \approx
\xi_{0,1}/( S \sqrt{\eta})$ as an estimation for the coherence length and  $Gi \approx
Gi_{1} S^2 \eta$ for the corresponding Ginzburg-Levanjuk parameter of the
two-band system, where $Gi_{1}$ is this quantity calculated separately for the
deep band. Thus, the enlargement of the inter-type domain in
Eq.~(\ref{eq:kappa_approx}) and the increase of $Gi$ is controlled  by the same
factor $S^2 \eta$. One concludes that a notable enlargement of the inter-type
domain can be achieved without compromising the validity of the mean field
calculations if $Gi_1$ is small enough. We also note that the above relations
suggest an interesting inverse dependence $\kappa_{li}^\ast - \kappa_{2}^\ast
\propto \xi_0^{-2}$ of the inter-type interval on the correlation length, or the
Cooper-pair size, $\xi_0$. 

In summary we predict that the inter-type domain between the two standard
superconductivity types is considerably enlarged in multi-band materials with a
shallow band when the latter yields a measurable contribution to the condensate
state. Thus for a wide range of microscopic parameters a multi-band
superconductor can fall into this domain. This will reveal itself in many
notable changes in the system's magnetic properties, which resemble the type I or
II superconductivity only in the close vicinity of the critical temperature
$T_c$. At lower temperatures such superconductors enter the inter-type
domain. Although a comprehensive description of the the inter-type domain has
not been achieved yet it is possible to predict that the mixed state in such
systems will exhibit  many unusual spatial vortex configurations not observed in
standard type II superconductors. 

\begin{acknowledgments}
The work was supported by Brazilian CNPq (grants 307552/2012-8 and 141911/2012-3) and FACEPE (grant 
APQ-0589-1.05/08). 
\end{acknowledgments}

\end{document}